\documentclass[12pt]{article}
\usepackage{aaspp4}
\usepackage{psfig}
\begin{document}

\title{ Temperature Variation in the Cluster of Galaxies Abell 115
Studied with ASCA}
 
\author{R. Shibata, H. Honda, M. Ishida}
\affil{Institute of Space and Astronautical Science, 3-1-1, Yoshinodai,
Sagamihara, Kanagawa 229-8510 Japan}
\authoremail{shibata@astro.isas.ac.jp}

\author{T. Ohashi}
\affil{Department of Physics, Tokyo Metropolitan University, 
1-1 Minami-Osawa, Hachioji, Tokyo 192-0397 Japan}

\author{K. Yamashita}
\affil{Department of Astrophysics, Faculty of Science, Nagoya University,
Furo-cho, Chikusa, Nagoya, Aichi, 464-8602, Japan}

\begin{abstract}

Abell 115 exhibits two distinct peaks in the surface brightness
distribution.  {\it ASCA} observation shows a significant temperature
variation in this cluster, confirmed by a hardness ratio analysis and
spectral fits.  A linking region between main and sub clusters shows a
high temperature compared with other regions. Two possibilities are
examined as the cause of the temperature variation: cooling flows in
the main cluster and a shock heating due to the collision of the
subcluster into the main system. Spectral fits with cooling flow
models to the main-cluster data show a mass-deposition rate less than
$419 M_{\odot}$ yr$^{-1}$.  Temperatures in the main cluster, the
linking region, and the subcluster are estimated by correcting for the
effects of X-ray telescope response as $4.9^{+0.7}_{-0.6}$,
$11^{+12}_{-4}$, and $5.2^{+1.4}_{-1.0}$ keV, respectively. The high
temperature in the linking region implies that Abell 115 is indeed a
merger system, with possible contribution from cooling flows on the
temperature structure.

\end{abstract}


\section{Introduction}

Clusters of galaxies are the largest gravitationally bound system in
the universe. They show us a large-scale distribution of matter in the
universe and help us understand the structure and evolution of
clusters themselves.  X-ray images of the hot, intracluster medium
(ICM) have been the most useful tool in determining cluster structures
and morphology, since the gas traces the cluster gravitational
potential generated by both visible and invisible matter.

Under a hierarchical structure formation scenario, substructures in
the ICM are formed when interaction or merging occurs between
subclusters. They disappear in a typical time scale (2 - 3 times the
sound crossing time, Roettiger {\it et al.}\ 1996, 1997) of a few Gyr,
therefore existence of a significant substructure strongly suggests
that a cluster is still evolving. A variety of morphology in clusters
has been revealed by the {\it Einstein} observatory.  A definite
scenario, however, does not exist concerning the detailed process of
the evolution of clusters.  For example, many clusters which have been
recognized as typical relaxed systems, are now known to have complex
temperature structures from {\it ASCA} (Honda {\it et al.}\ 1996). This
clearly indicates that it is not enough to judge the evolutionary
stage of clusters only from the surface brightness distribution, but
we need to examine distributions of ICM temperature and metal
abundance.

The cluster of galaxies Abell 115 (A115, Richness class = 3) is a
fairly distant cluster ($z$ = 0.1971, Distance class = 6) and known in
the literature to be characterized by a double peak X-ray surface
brightness (Bautz-Morgan type = III, Abell {\it et al.}\ 1989).  This
morphology suggests that A115 is possibly in the process of a
merger.  The brightest galaxy in the center of the X-ray peak is
associated with a radio source 3C 28.

The strong radio source 3C 28 was observed at 4.9 GHz with the Very
Large Array (VLA) in 1984 April (Giovannini {\it et al.}\ 1987).  High
resolution and high sensitivity radio maps show two extended
components located on northern and southern sides of the optical
galaxy.

In the optical band, redshifts for 19 galaxies which are certainly
associated with A115 were measured with the Kitt Peak National
Observatory (KPNO) 4 meter telescope (R-band) (Beers {\it et al.}\ 
1983).  These plates reveal the presence of three primary clumps of
galaxies.  The two peaks in the X-ray surface brightness map
correspond to the two concentrations in the galaxy distribution in the
plates.  No X-ray emission is, however, detected from the third clump
which is $7'$ east of the X-ray main peak by the {\it Einstein
observatory}.

X-ray observations were performed with the Imaging Proportional
Counter (IPC) in January 1979 (2.6 ksec) and the High Resolution
Imager (HRI) in February 1981 (8.2 ksec) on board the {\it Einstein
Observatory}. The IPC image revealed its X-ray surface brightness
distribution with two clear enhancements (Forman {\it et al.}\
1981). The {\it ROSAT} HRI observation was carried out in January and
July 1994 (21 ksec and 30 ksec, respectively).  The X-ray image in the
archival data shows a similar double structure.  The {\it Einstein}
HRI observation showed no point-like source in the center of the main
X-ray peak, and an upper limit to a point-source flux was estimated as
$\lesssim 1.5 \times 10^{-13}$ erg cm$^{-2}$ s$^{-1}$ (Feretti {\it et
al.}\ 1984).  Based on an image deprojection analysis of the {\it
Einstein} imaging data, the main peak in A115 indicates a large
cooling flow with a mass-deposition rate $\dot{M} = 313^{+369}_{-186}
M_{\odot}$ yr$^{-1}$ (White {\it et al.}\ 1997).

Previous X-ray observations, however, did not show detailed properties
of the ICM in A115, except for its morphology.  To examine
dynamical state of the cluster, we need to look into structures of
temperature and metal abundances with enough sensitivity.
In this paper, we present results on the ICM properties of A115
obtained with {\it ASCA}\@. The large effective area covering up to 10
keV and the high spectroscopic resolution of {\it ASCA} (Tanaka {\it
et al.}\ 1994) are particularly useful for the study of this cluster.
Throughout this paper, we adopt $H_0 = 50$ km s$^{-1}$Mpc$^{-1}$ and
$q_0= 0$. At the distance of A115, $5'$ corresponds to about 1.7
Mpc.

\section{ {\it ASCA} Observation}\label{sec-obs-log}

{\it ASCA} observation of A115 was carried out from 1994 Aug 5.96
to 7.02 (UT). The satellite carries four identical grazing-incidence
X-Ray Telescopes (XRT, Serlemitsos {\it et al.}\ 1995, Tsusaka {\it et
al.} 1995).  Each XRT has approximately $3'$ half power beam
width. The point spread function (PSF) of the XRT has a cusp-shaped
peak. This property allows us to partly resolve small-scale structures
in clusters of galaxies. Two Solid-state Imaging Spectrometers (SIS0
and SIS1, Burke {\it et al.}\ 1991) and two Gas Imaging Spectrometers
(GIS2 and GIS3, Ohashi {\it et al.}\ 1996, Makishima {\it et al.}\
1996) are placed at the focal plane of the four XRTs.  SIS has a Field
Of View (FOV) of $22'\times 22'$ and covers an energy range 0.4 to 10
keV with a spectral resolution of 180 eV (FWHM) at 6 keV\@.  GIS has a
FOV of $50'$ diameter and its energy range is from 0.7 to 10 keV\@.
This instrument has a higher detection efficiency than SIS at the
upper end of the energy band. These characteristics of {\it ASCA}
allow us a sensitive study on the spatial structures of temperature
and metal abundance in clusters of galaxies.

Throughout the observation of A115, GIS and SIS were operated in
the normal PH mode and in the 1CCD Faint mode (FOV is $11'\times
11'$), respectively. We screened the data by requiring the cut-off
rigidity to be $> 6$ GeV $c^{-1}$ and $> 8$ GeV $c^{-1}$, for GIS and
SIS, respectively, and the target elevation angle above the earth rim
to be $> 5^{\circ}$.  In the SIS data selection we used event grades
of 0, 2, 3, and 4, and further required the elevation angle above the
day earth rim to be $> 25^{\circ}$.  We also discarded data when the
spacecraft was in high background regions such as the South Atlantic
Anomaly (SAA)\@.  After removing data within $\sim 1.2$ hr  just
after the maneuver to A115, fluctuation of the pointing direction
was kept within about $1'$ in radius.
We then added all of the available SIS data from the 2 sensor units
(SIS 0 and SIS 1), regardless of the chip selection and data modes,
and produced a single SIS spectrum after an appropriate gain
correction.  Similarly, data from GIS 2 and GIS 3 were summed into a
single GIS spectrum.

Background (BGD) spectra were generated from blank-sky data in the
public archive.  Photons in the same detector region were accumulated
for the source and for BGD\@.  The blank sky data of SIS is taken only
in 4CCD mode.  Since the level of non X-ray background (NXB) is
different between 4CCD and 1CCD modes, we adjusted it by normalizing
with the detected intensity of Ni K$_{\alpha}$ and K$_{\beta}$ lines,
which are excited by the NXB flux.

Table \ref{table_1} summarizes exposure times after the data
selection, counting rates and fluxes, calculated for single
temperature Raymond-Smith models (Raymond \& Smith 1977) assuming $kT
= 5$ keV, for the whole cluster after the BGD subtraction.

\placetable{table_1}

\section{Results}
\subsection{Image analysis}\label{sec-image-ana}

\begin{figure}[htb]
\begin{center}
\mbox{\psfig{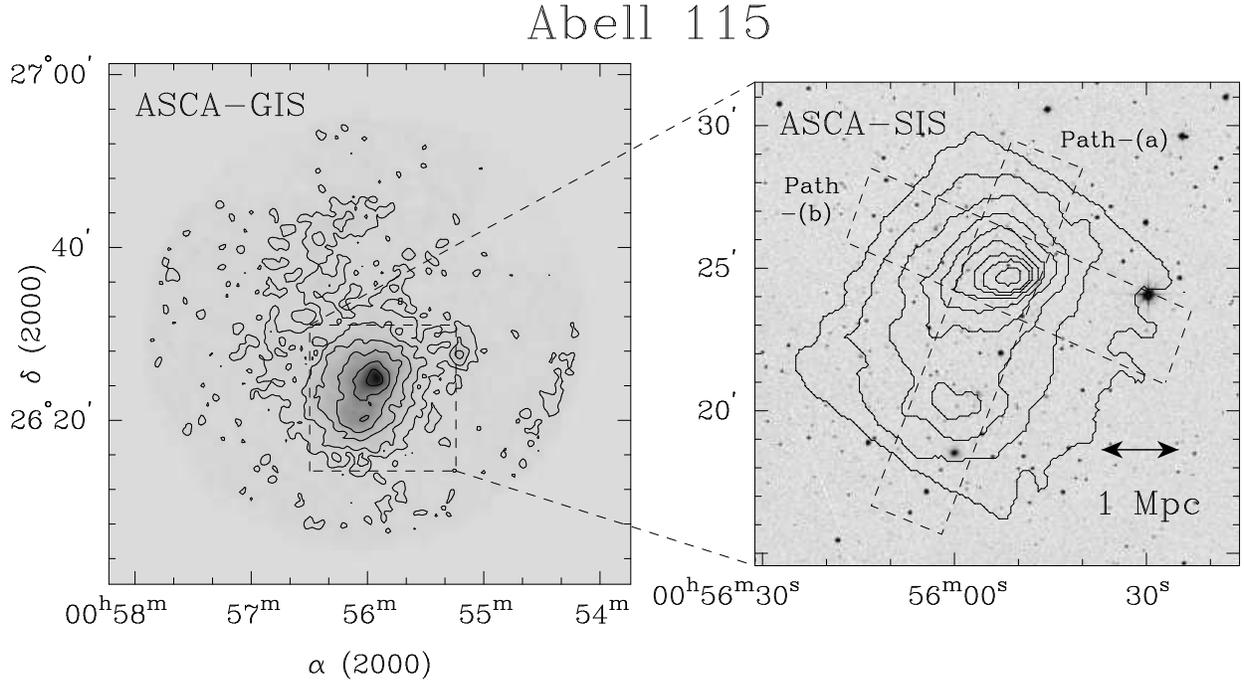}}
\caption{Left panel shows a greyscale GIS image of A115, superimposed
on isointensity contours in a $60' \times 60'$ field centered on the
field direction of GIS. Contour levels are logarithmic, showing 2.00,
4.15, 8.63, 17.9, 37.3, and 77.6 times the BGD level.  Right panel
also shows the Digitized Sky Survey image overlaid on a contour plot
of the SIS image. The lowest contour level roughly corresponds to the
edge of the field of view of SIS\@.  Contour levels are 8.7, 11.6,
15.5, 20.6, 27.4, 36.5, 47.6, 64.7, and 86.3 times the BGD level. Two
dashed rectangles indicate regions where data for the hardness ratio
analysis are accumulated (see Figure 2).  In both images, background
images are not subtracted.}
\label{figure_1}
\end{center}
\end{figure}

Figure \ref{figure_1} shows the surface brightness distribution of
A115 observed with {\it ASCA}. These images are smoothed by a
Gaussian function with $\sigma = 0.'32$. Left figure shows a greyscale
{\it ASCA} GIS image, superposed on isointensity contours, in a $60'
\times 60'$ field centered on the FOV\@. Right figure also shows the
Digitized Sky Survey (DSS) image overlaid on the contours of {\it
ASCA} SIS image. The lowest brightness contour roughly corresponds to
the edge of the SIS FOV\@.  Both contour levels are logarithmically
evenly spaced and background image is not subtracted.
\placefigure{figure_1}
In figure \ref{figure_1}, the main peak (centered at R.A.[J2000] =
$0^h 55^m 55^s$, Dec.[J2000] = $26^{\circ}24' 48''$) and the sub peak
(centered at R.A.[J2000] = $0^h 56^m 04^s$, Dec.[J2000] =
$26^{\circ}20' 33''$) are separated clearly, and no optical sources in
the DSS image exist at the position of the X-ray main peak.  
We can confirm that A115 has an X-ray extent of more than $15'$
($\sim 5$ Mpc ) in diameter, and the surface brightness distribution
obtained with {\it ASCA} is similar to that of {\it Einstein} IPC
(Forman {\it et al.}\  1981).

In the image analysis, we first fit the SIS radial profile with the
$\beta$ model.
The $\beta$ model surface brightness $I_x(r)$ at a projected radius $r$ 
varies as 
\[
I_x(r) = I_0 \left \{ 1 + \left ( \frac{r}{R_c} \right ) ^2 \right \}
^{-3 \beta + 1/2} ,
\]
where $R_c$ is the core radius and $I_0$ is the central brightness.

To obtain the $\beta$ model parameters which describe the X-ray
emission from A115, we produced brightness distribution maps for
various $\beta$ models and folded them with the PSF of the XRT\@.
These simulated profiles were then compared with the SIS radial
profile.  We adjusted the parameters until the simulated profile gave
a consistent fit to the observed profile.
The parameter values for the main peak are $\beta =
1.05^{+0.08}_{-0.06}$ and $R_c = 1'.22^{+0'.29}_{-0'.24}$,
respectively, giving a $\chi^2_{\nu}$ value of 1.14 (d.o.f. = 55).  We
attempted to fit the radial profile for the sub peak. Due to the poor
statistics, however, we could not constrain either $\beta$ or $R_c$
value. We thus fixed both $\beta$ and $R_c$ at the best-fit values for
the main peak, and obtained normalization to be $0.33 \pm 0.03$ times
that of the main peak.  The errors indicate 90\% statistical errors
for single parameter. With these $\beta$ model parameters, the
observed brightness distribution is well reproduced.

\subsection{Spatial variation of the hardness ratio}

\begin{figure}[htb]
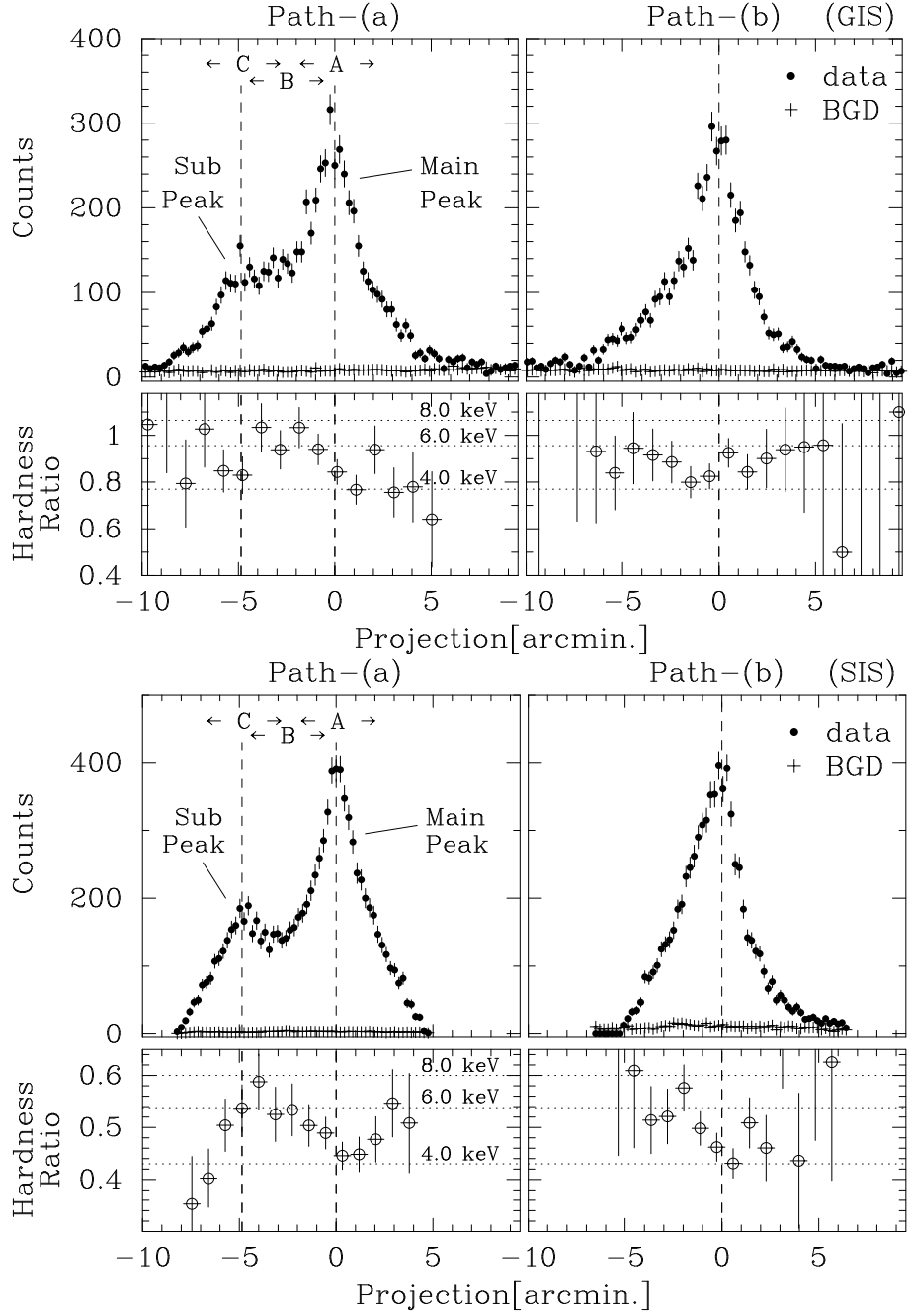

\begin{center}
\mbox{\psfig{figure=figure_2gis.ps,width=0.73\textwidth,angle=270}}\\
\mbox{\psfig{figure=figure_2sis.ps,width=0.73\textwidth,angle=270}}
\caption{Intensity and hardness ratio profiles of A115 along 2 paths
measured with GIS and SIS.  Left and right panels are for path-(a) and
path-(b) accumulated in a rectangular region having a width of $2.7'$
(see figure \ref{figure_1}).  Top and bottom panels are for the GIS
and the SIS data, and filled circles and crosses indicate raw and
background data, respectively.  The arrows also indicate the regions
A, B, and C with $2'$ radius each, which are used for the spectral
analysis in section \ref{sec-spec-all}.  Bottom diagram in each panel
shows hardness ratio profile for counting rates between $2 - 10$ keV
and $0.5 - 2$ keV\@.  The dotted lines indicate corresponding
temperatures for Raymond-Smith models (4.0, 6.0, and 8.0 keV).}
\label{figure_2}
\end{center}
\end{figure}

Figure \ref{figure_2} shows cross sections of the SIS and the GIS
images in figure \ref{figure_1}, along the path connecting the two
emission peaks (Path-(a) in figure \ref{figure_2}) and its vertical
direction (Path-(b) in figure \ref{figure_2}).  The angular width of
the data accumulation is $2.7'$ for each path.  The bottom panels show
Hardness Ratios (H.R.) of the counting rates between the energy bands
$2 - 10$ keV and $0.5 - 2$ keV\@.  The dotted lines indicate
corresponding temperatures for single temperature Raymond-Smith
models.  The H.R. for path-(a) shows the maximum between the main and
the sub peaks, and becomes lower toward the main peak.
\placefigure{figure_2}
If we fit this H.R. variation with a constant value, the
$\chi^2_{\nu}$ value becomes 1.83 (d.o.f. = 6), indicating that a
significant variation in the H.R. exists with a confidence level of
more than 90\% (Fig. 2; Path-(a)(SIS)).  This is significant even if
we allow the BGD level to vary by $\pm10\%$. Therefore, the
H.R. distributions in figure \ref{figure_2} indicate a real
temperature variation in the ICM\@.
The corresponding temperatures are $6 \sim 8$ keV at the sub peak and
4 keV in the main cluster, respectively, if we assume the single
temperature thermal emission.

The H.R. along the path-(b) is low near the main peak, similarly to
that in the path-(a). Thus the central region of the cluster main peak
may have a lower temperature than the outer regions.  Because of the
limited statistics, however, we can not definitely conclude from these
results whether there exists an additional cool component such as the
cooling flows suggested from the {\it Einstein} deprojection analysis
(White {\it et al.}\ 1997).

\subsection{Spectral analysis}\label{sec-spec-all}

\begin{figure}[htb]
\begin{center}
\mbox{\psfig{figure=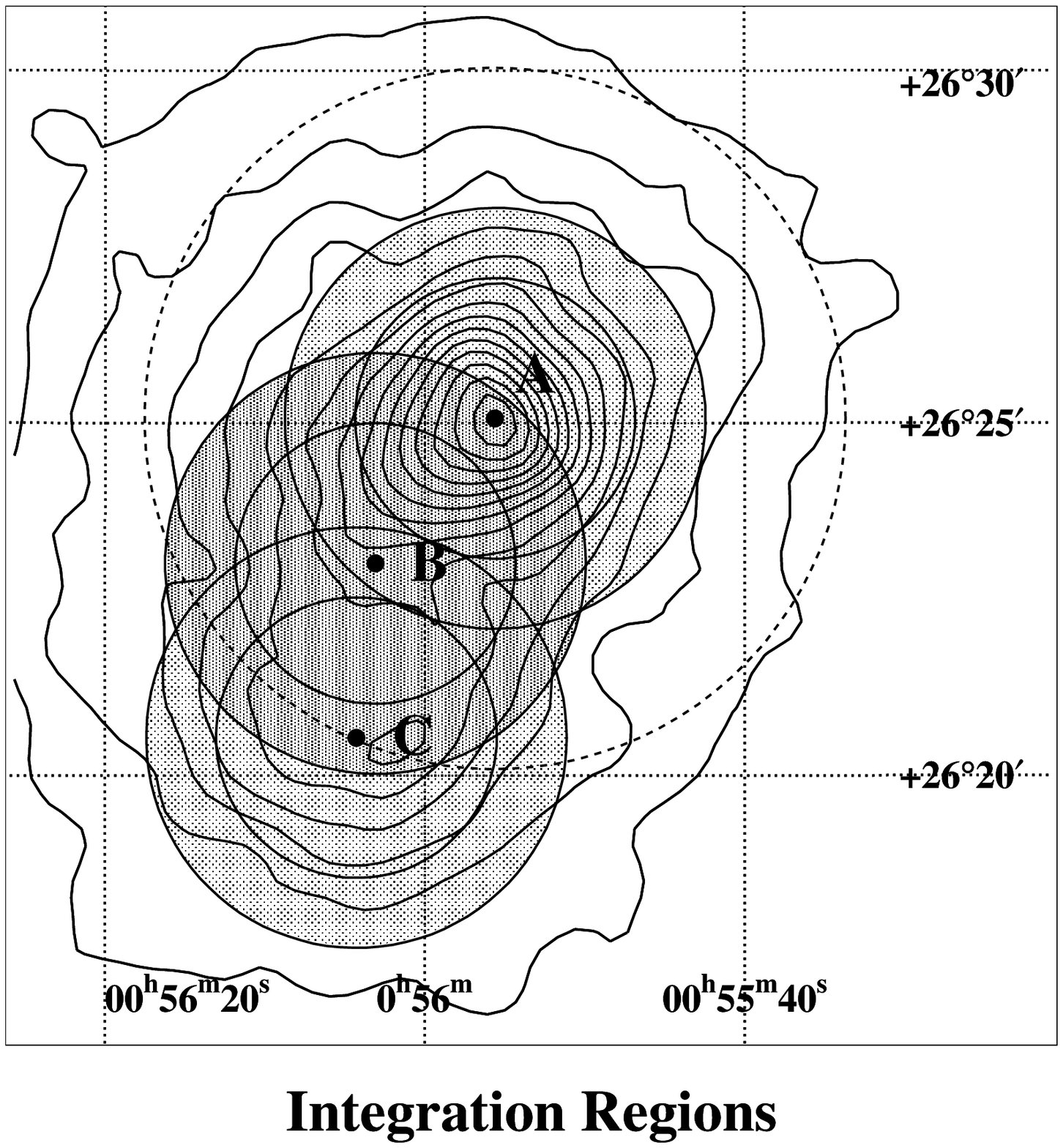,width=0.99\textwidth,angle=0}}
\caption{Grey-scale plot of integration regions for the spectral
analysis overlaid on the GIS image.  Spectra in these seven
integration regions are analyzed in section \ref{sec-spec-all}.  The
centers of the regions correspond to the main peak (region A), sub
peak (region C) and the midpoint between the two peaks (region B),
respectively.  The circles with slid lines indicate radii of $2'$ and
$3'$, and spectral analyses are carried out for GIS and SIS data.  The
broken circle with a radius of $5'$ indicates a region for GIS spectral
analysis. }
\label{figure_3}
\end{center}
\end{figure}

Following the hardness ratio analysis, we investigate the variation of
energy spectrum over the entire cluster.  The pulse-height spectra are
accumulated in 3 circular regions with a radius of $2'$ or $3'$.  One
centered on the main peak (northeast direction, R.A. = $0^h 55^m
55^s$, Dec.\ = $26^{\circ}24' 48''$, hereafter region A), one at the
sub peak (south direction, R.A. = $0^h 56^m 04^s$, Dec.\ =
$26^{\circ}20' 33''$, hereafter region C) , and one in an intermediate
point between the two peaks (center of A115, R.A. = $0^h 56^m 03^s$,
Dec.\ = $26^{\circ}23' 01''$, hereafter region B). As for the GIS data,
larger integration radii of $5'$ and $10'$ centered on the main peak
are also used. Figure \ref{figure_3} shows these integration regions
overlaid on the GIS image.
\placefigure{figure_3}

In calculating the response function for the spectral fit, we assume
that the surface brightness of the cluster is described by a double
$\beta$ model as derived in section \ref{sec-image-ana}. The
parameters are $\beta = 1.05, R_c = 1'.22$, and the ratio of
normalization between the 2 peaks = 1 : 0.33.  This response also
assumes that the cluster is isothermal. This assumption differs from
the actual case, but it enables us to perform spectral fits separately
in individual regions and gives approximate temperatures (see Honda {\it et
al.}\ 1996). This method enables us to detect temperature variations in
clusters, as already demonstrated for the Coma cluster (Honda {\it et
al.}\ 1996, Kikuchi {\it et al.}\ 1998).

\begin{figure}[htb]
\begin{center}
\mbox{\psfig{figure=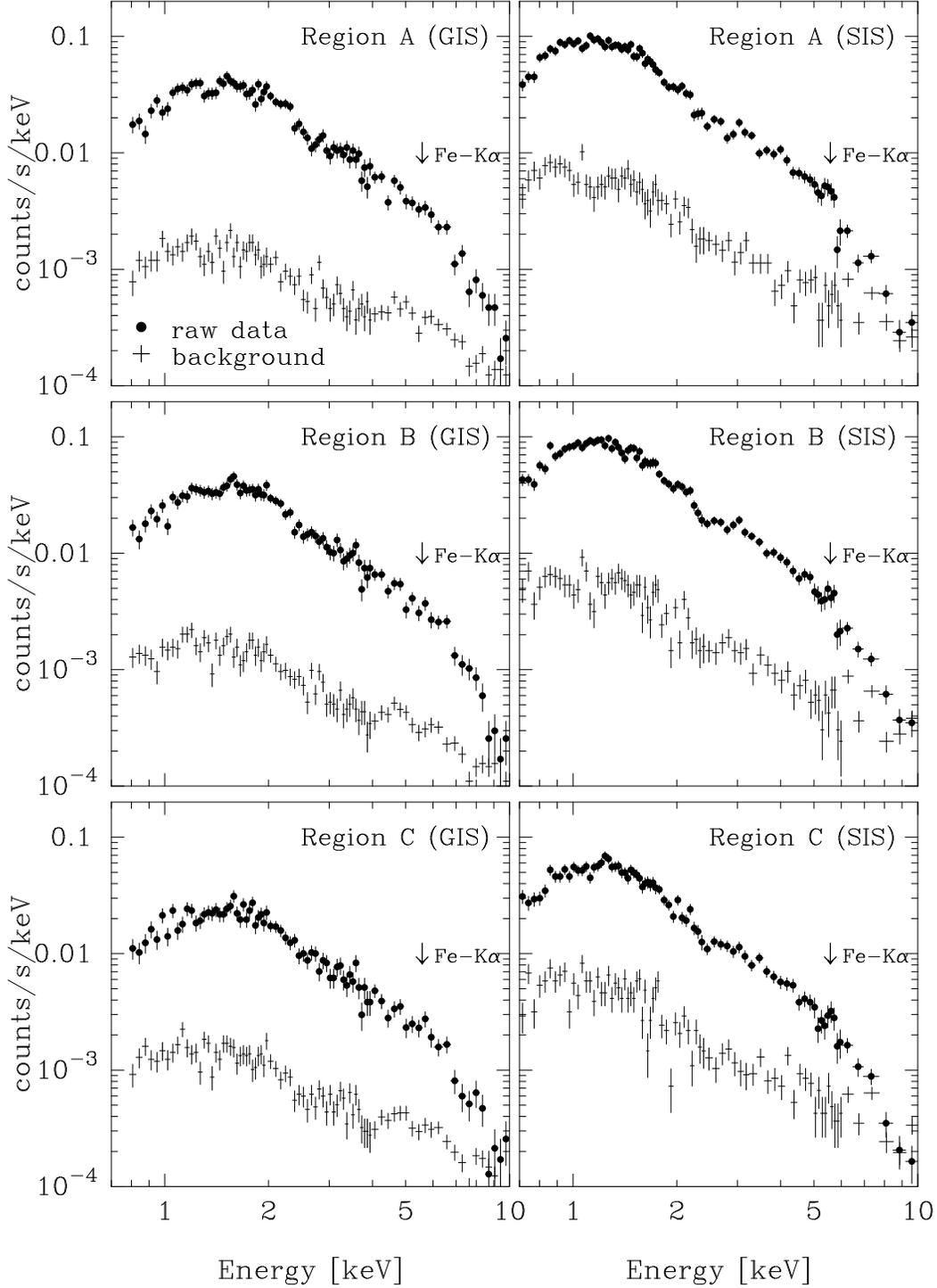,width=0.90\textwidth,angle=0}}
\caption{Observed GIS and SIS spectra integrated in a radius of
$3'$. Left and right panels correspond to GIS and SIS,
respectively. Top, middle, and bottom panels are for regions A, B, and
C, respectively.  Filled circles and crosses indicate raw and
background spectra, and arrows show expected energy for a Fe
K$_{\alpha}$ emission line.  }
\label{figure_4}
\end{center}
\end{figure}

Figure \ref{figure_4} shows the observed GIS and SIS spectra for the
on-source and the background data before subtracting the background.
The background (CXB + NXB) level is less than 10\% of the observed
flux. Table \ref{table_2} summarizes the number of photons in each
integration region.
\placefigure{figure_4}
\placetable{table_2}

\begin{figure}[htb]
\begin{center}
\mbox{\psfig{figure=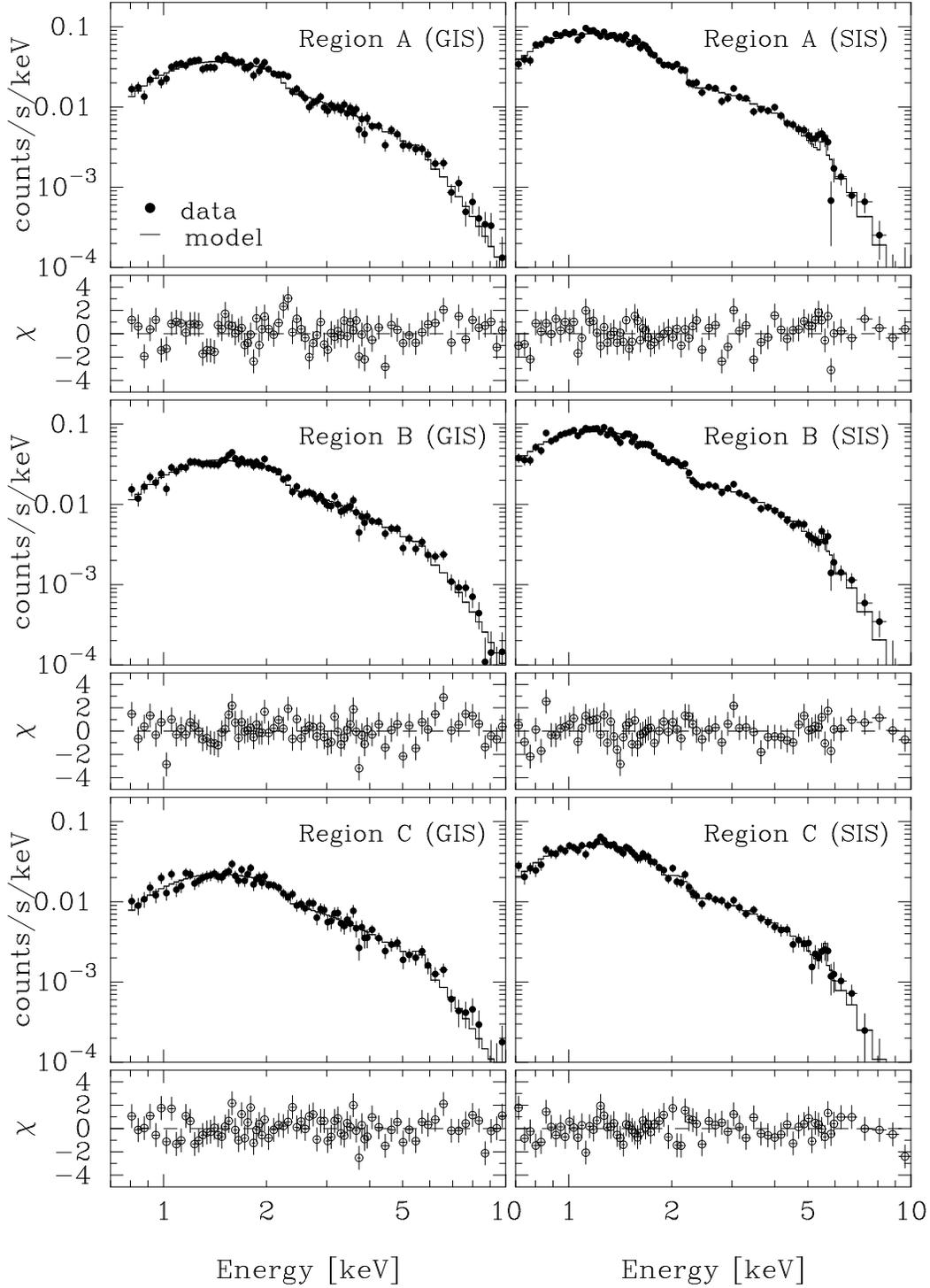,width=0.90\textwidth,angle=0}}
\caption{The background subtracted GIS and SIS spectra fitted thermal
models. The integrated radius is $3'$. The layout is the same as that
in figure \ref{figure_4}.  Filled circles, lines, and open circles
indicate observed data, best-fit models, and residuals of the fit,
respectively. We adopted single temperature Raymond-Smith models with
a redshift $z = 0.1971$.}
\label{figure_5}
\end{center}
\end{figure}

Figure \ref{figure_5} shows results of the spectral fit.  Table
\ref{table_3} summarizes the spectral parameters for each integration
region.
\placefigure{figure_5}
\placetable{table_3}
The spectral fits are performed in the energy band 0.7 - 10 keV and
0.8 - 10 keV for SIS and GIS, respectively.  Raymond-Smith and
power-law models at a redshift $z = 0.1971$ are assumed. Except for
the region C which have very poor statistics, the power-law models are
excluded. The best-fit galactic column densities turn out to be
roughly 50\% larger than the published value of $5.1 \times 10^{20}$
cm$^{-2}$ (Stark {\it et al.}\ 1992).  This is partly due to the
uncertainty in the SIS response in the low energy band, and has no
significant influence on other spectral parameters.
For the integration regions A, B, and C, the BGD contributions are
7\%, 9\%, and 11\% in a radius of $2'$, respectively. The derived ICM
temperature in each region varies by about 0.1 keV when the BGD
normalization is varied by $\pm10\%$. Uncertainty in the BGD level do
have some effect on the energy response function through the change of
the $\beta$ model parameters. Its effect on the temperature is,
however, 0.05 keV or less for all regions.

The temperature for a region with $3'$ radius is higher than that with
$2'$ for the regions A (main peak) and C (sub peak).  On the other
hand, integration over a $2'$ radius indicates higher temperature than
$3'$ radius in the region B (linking region). This implies that the
temperature takes the maximum value somewhere between the main and the
sub peaks, and decreases toward each X-ray peak.  This feature is
consistent with the result of the H.R. analysis shown in figure
\ref{figure_2}.

More than half of the photons falling in region B ( $r = 2', kT =
6.25$ keV) comes from region A ( $r = 2', kT = 4.85$ keV) (figure
\ref{figure_7}, section \ref{morekomi}), and the fraction of photons
originating from the corresponding sky region is only 40\%. The fact
that the spectral fit for the region B data gives the highest
temperature ($\sim 6$ keV) indicates that the true temperature in this
region is even higher.

\section{Origin of the temperature variation}

Both the H.R. variation shown in figure \ref{figure_2} and the
spectral analysis in section \ref{sec-spec-all} indicate that the ICM
temperature is high in the linking region between the main and the sub
peaks.  We will look into the possible processes which can cause the
observed temperature variation along 2 broad scenarios: a cooling flow
in the main cluster and a shock heating caused by a subcluster merger.

\subsection{Cooling flows}\label{sec-spec-main}

If the center of the main peak has an intense cool component (cooling
flows), then only the outer regions show the virial temperature of the
cluster. The angular response of the XRT would cause the the cool
component spread out in the detector plane over a few arcminutes. The
H.R. profile in figure \ref{figure_2} in fact indicates that the
central $3'-4'$ of the main peak has a soft spectrum, although the SIS
profile along path-(a) is asymmetric around the main peak.  

The {\it Einstein} observation showed that the main peak of A115
had the large mass-flow rate $\dot{M} = 313^{+369}_{-186} M_{\odot}$
yr$^{-1}$ (White {\it et al.}\ 1997).  The X-ray luminosity of the
cooling flow with this mass-flow rate is estimated to be
\[
L_{\rm cool}  \simeq  7.9 \times 10^{43} 
\left (\frac{kT_{\rm cool}}{1\ {\rm[ keV]}} \right ) 
\left ( \frac{\dot{M}}{313\ [M_{\odot} {\rm yr}^{-1}]} \right ) 
\ {\rm [erg\ s}^{-1}]  ,
\]
where $kT_{\rm cool}$ is an assumed temperature of the cool component.
At the distance of A115, the expected flux is $F_{\rm cool}
\simeq 4.7 \times 10^{-13}$ erg cm$^2$ s$^{-1}$.

If the cooling flow occurs within 100 kpc ($\simeq 0.3'$) of the
center of the main peak, $L_{\rm cool}$ contributes about 16\% of the
flux in the energy band 0.5 - 10 keV in the region A with $r = 2'$.
Table \ref{table_cf} summarizes results of spectral fits with various
cooling flow models (Mushotzky \& Szymkowiak, 1988).  Because the
mass-flow rate tightly couples with the galactic column density in
this spectral fit, we fix it to the best fit value obtained in the
previous spectral analysis (section \ref{sec-spec-all}). 

The spectral fit was unable to constrain the slope parameter $s$ of
the emission measure (EM) against temperature, when $s$ was a free
parameter. We examined the spectrum by fixing the $s$ values between
$-0.5$ and $+0.7$.  This range corresponds to the 90\% error when the
mass-flow rate is set to $313 M_{\odot}$ yr$^{-1}$ and includes the
best-fit values ($s \sim 0.7$) for many other clusters reported by
Mushotzky \& Szymkowiak (1988).  As shown in table \ref{table_cf} ,
the observed spectra can be fitted with a cooling flow model with the
mass-flow rate reported by {\it Einstein}.  In this case, the ICM
temperature in region A becomes $5.5 \sim 6.5$ keV, which is
consistent with that in region B measured within $r = 2'$.  This
result indicates that we can interpret the temperature variation in
A115 in terms of the cooling flow picture; i.e.\ the virial
temperature in the main peak is in reality $\sim 6$ keV, which is
suppressed in the ASCA data due to the presence of a cooling flow, and
other regions in the cluster are filled with this $\sim 6$ keV ICM\@.
We can then estimate confidence limits for the mass-flow rate based on
the {\it ASCA} spectra for various fixed values of $s$. An upper limit
for $\dot{M}$ is obtained as $419 M_{\odot}$ yr$^{-1}$ in the case of
$s = 0.58$, which is lower than the limit set by
the {\it Einstein} IPC (White {\it et al.}\ 1997).

\begin{figure}[htb]
\begin{center}
\mbox{\psfig{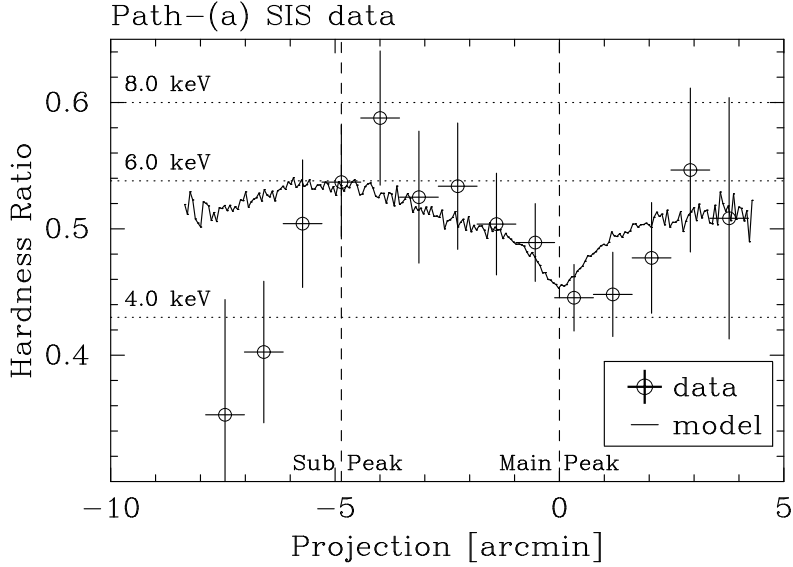}}
\vspace{-4mm}
\caption{The simulated H.R. profile assuming a
 cooling flow with $\dot{M} =313 M_{\odot}$ yr$^{-1}$ exist in the main
peak.  Solid line indicates the simulation result and crosses show
observed data.  }
\label{figure_cf_hr}
\end{center}
\end{figure}

We then calculated the expected H.R. profile assuming the
cooling flow rate of $\dot{M} =313 M_{\odot}$ yr$^{-1}$ in the main
peak.  Assuming that the main ICM has a temperature of 6 keV with the
spatial extent as obtained in section \ref{sec-image-ana} and the cool
1 keV component with an extent of 100 kpc ($\simeq 0.3'$), the SIS
image has been simulated because it has better spatial resolution than
the GIS\@.  Figure \ref{figure_cf_hr} shows the calculated profile
overlaid on the H.R. data (Path-(a)(SIS) in figure \ref{figure_2}).  The
cool region has a radius of about $3'$ due to the PSF of the
telescope.  $\chi^2_{\nu}$ value for the simulated profile with the
actual data is 1.0 (d.o.f.\ = 14), and the two profiles look similar
with each other. Therefore, the cooling flow reported by the {\it
Einstein} observation can account for the spatial variation of the
temperature in A115\@.

\subsection{Merger}\label{morekomi}\label{sec-real-temp}

The spectral fit of the main peak data (region A, $R=2'$) with a
cooling flow model showed the best-fit mass-flow rate to be $89
M_{\odot}$ yr$^{-1}$, which is much smaller than the {\it Einstein}
result. The {\it ASCA} spectrum is represented very well by the single
temperature Raymond-Smith model (table \ref{table_cf}).  Regarding the
fact that the highest temperature appears at the linking region
between the main and the sub peaks, we should also interpret this
feature in terms of shock heating due to a collision of the two
subclusters. In this section, we will estimate the correct
temperatures in the 3 regions in the cluster.
%and by taking into account contaminations from outside the field of view

Since the PSF of {\it ASCA} XRT has an extended tail and the integrated
regions A, B, and C are not far apart, significant fraction of photons
detected in each region is contaminated by those from other sky
regions.  Therefore, we first estimate fractions of photon origins for
each integration region.

\begin{figure}[htb]
\begin{center}
\mbox{\psfig{figure=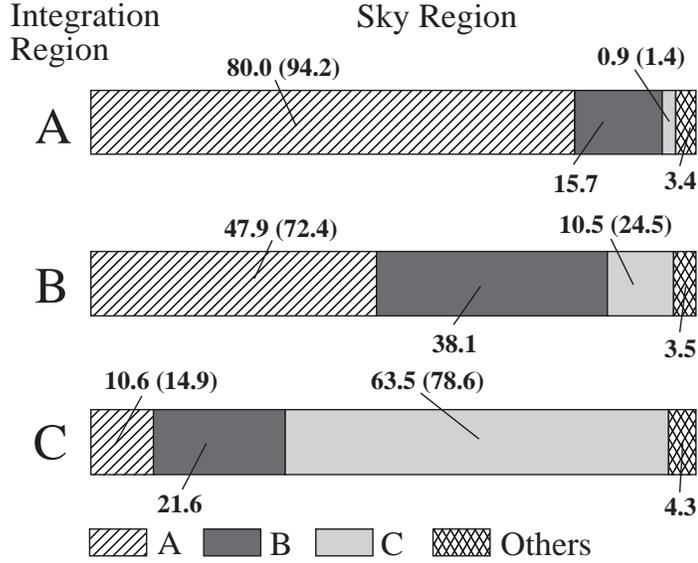,width=0.63\textwidth,angle=0}}
\vspace{-4mm}
\caption{Fractions for the origins of photons detected in each
integration region with $2'$ radius (SIS data).  Because of the tail
of the point spread function, photons in each integration region are
contaminated by those originating from the other sky regions.  The
three integration regions overlap with each other, and the numbers
indicate the fractions when photons from the overlapped region are
assigned as the region B\@.  In round brackets, we also indicate the
values when the data in the overlapping regions are assigned to either
A or B.  }
\label{figure_7}
\end{center}
\end{figure}

The surface brightness distribution of A115 is approximated by a
double $\beta$ model with the parameters $\beta = 1.05, R_c = 1'.22$,
and the flux ratio of the 2 peaks is 0.33.  We then calculate, using
the {\it ASCA} simulator, an expected image when photons of this
surface-brightness distribution go through the XRT and reach the
detectors.  We then trace back each photon to the original sky
coordinate.  Figure \ref{figure_7} shows fractions of 
photon origins for each integration region thus estimated.
\placefigure{figure_7}
As shown in this figure, 94\% and 79\% of the photons collected in the
regions A and C, respectively, come from corresponding sky
regions. There is, therefore, little influence from surrounding sky
regions on the spectral results described in section
\ref{sec-spec-all}.  On the other hand, region B data contain only
38\% of the photons produced in the corresponding sky, and a high
fraction of photons come from region A ($\sim 48\%$) and C ($\sim
11\%$). The measured temperature for region B is then roughly an
weighted average of the values in regions A, B and C.

To obtain the true temperatures of the sky regions, influence from
other sky regions needs to be subtracted.  This can be done using a
ray-tracing simulation. True temperatures are estimated for regions A,
B and C within $r < 2'$ using the SIS data, which are less
contaminated because of the better position resolution.  The actual
process is an iterative one as described below.

\begin{description}

\item[(a)] To obtain a better estimate of the temperature in region C,
calculate a contaminating spectrum from region A to C using the XRT
and SIS responses and assuming the temperature of region A to be 4.97
keV\@. Then subtract it from the observed SIS data for the region C,
and evaluate the temperature with the spectral fit. Note that the
contamination from region B to C is negligible.

\item[(b)] Calculate a contaminating spectrum from the regions A ( $kT
 = 4.97$ keV ) and C ( with the temperature determined in step (a) )
 to B\@. Then subtract it from the observed spectrum B, and evaluate
 its temperature.

\item[(c)] In the same way as in step (b), simulate the spectrum from
the regions B and C to A\@.  Subtract it from the observed
spectrum A, and evaluate its temperature.

\item[(d)] Repeat the process (a) through (c) until the changes of
temperatures of all the regions become less than 0.01 keV\@.

\end{description}

We assumed that the surface brightness follows 2 $\beta$ models for
the main and the sub peaks and calculated the response function for
the spectral fit. The integration regions A, B, and C, all having a
$2'$ radius, overlap with each other, and we allocated these
overlapping data into region B in the spectral analysis.
Table \ref{table_4} shows the corrected temperatures compared with those
evaluated in the previous section, only for the SIS data.
\placetable{table_4}
As a result, the realistic temperatures of the three sky regions
become $kT_A = 4.9 ^{+0.7}_{-0.6}$ keV (main peak), $kT_B = 11
^{+12}_{-4}$ keV (between both peaks), and $kT_C = 5.2 ^{+1.4}_{-1.0}$
keV (sub peak), respectively.  The errors are evaluated by fixing the
temperatures in other regions at their best-fit values. If temperature
$kT_A$ is varied within this error range, the $kT_B$ value changes by
$^{+2.0}_{-1.5}$ keV\@.

This analysis shows that the temperature in region B is significantly
higher than that in region A\@.  As seen in table \ref{table_4}, the
regions A and C maintain almost the same temperature as the previous
values in section \ref{sec-spec-all}.  In contrast, fairly large
change of temperature occurs in B\@. As shown in figure
\ref{figure_3}, the edge of region B is close to the main and the sub
peaks of the surface brightness which have lower temperatures, even
taking $r = 2'$.  If we were able to limit the data within $r = 1'$,
the temperature in this region would have become even higher; although
it is not feasible because of too large statistical errors. To
conclude, we obtain a strong suggestion that the ICM temperature in
region B is very high.

\section{Discussion}

{\it ASCA} observations of A115 have revealed significant
temperature structures in the ICM\@.  The H.R. variation along the
path connecting the two intensity peaks (Figure \ref{figure_2}
Path-(a)) indicates that the temperature between the main and the sub
peak is the highest. The spectral analysis based on an approximated
response confirms the same feature (section \ref{sec-spec-all}).  The
cooling flow reported by the {\it Einstein} observation in the main
cluster can account for this temperature variation (section
\ref{sec-spec-main}).  The mass-flow rate is obtained as $89
M_{\odot}$ yr$^{-1}$ with an upper limit of $419 M_{\odot}$ yr$^{-1}$,
whose values are smaller than $313^{+369}_{-186} M_{\odot}$ yr$^{-1}$
reported by {\it Einstein}. The shock heating triggered by the merging
is another possible interpretation for this feature, as suggested by
the cluster morphology (section \ref{morekomi}).

In the case of Hydra A cluster, which has been recognized as a typical
cooling flow cluster with a very large mass-flow rate ($\dot{M} \sim
600 M_{\odot}$ yr$^{-1}$, David {\it et al.}\ 1990) from the {\it
Einstein} data, {\it ASCA} observation indicated a mass-flow rate of
only $\dot{M} \sim 60 M_{\odot}$ yr$^{-1}$ (Ikebe {\it et al.}\ 1997).
This is an order of magnitude smaller than the previous estimate.  The
spectral information in the harder energy band, covering the hot
component emission, is necessary for the correct estimation of the
cool component flux.

We can estimate the central gas density $n_0$ and the cooling time in
the main cluster as follows.  The density $n_0$ is evaluated by taking
$L_X \sim 9.1 \times 10^{44}$ erg s$^{-1}$ and a $\beta$ model gas
distribution in a volume $V \sim 10$ Mpc$^3$ in the within $r < 2'$ in
region A\@. This gives $n_0 \simeq 3.3 \times 10^{-3}$ cm$^{-3}$:
which is comparable to the {\it Einstein} IPC result by Forman {\it et
al.}\ (1981) of $1.8 \times 10^{-3}$ cm$^{-3}$, but considerably lower
than that estimated by Feretti {\it et al.}\ (1984) from the {\it
Einstein} HRI data ($n_0 = 4.7 \times 10^{-2}$ cm$^{-3}$).  We regard
that the HRI result is rather weighted on the interstellar matter
associated with 3C28 because of the very small core radius (35 kpc).
Taking the central densities by {\it ASCA}, cooling time is estimated
to be $1.7 \times 10^{10}$ yr which is comparable to the Hubble time.
Though we have some uncertainty in the central density and the volume
of ICM, it seems difficult for the cooling flow with a high $\dot{M}$
to develop in the center of A115 (around the radio galaxy 3C28).

On the other hand, we find it also possible to interpret the high
temperature region in terms of a collision between the two
subclusters.  After correcting for the effects of contaminating
photons from outside of the integrated sky regions (section
\ref{sec-real-temp}), we evaluated the temperatures in three regions
with $2'$ radius to be $kT_A = 4.9^{+0.7}_{-0.6}$ keV, $kT_B =
11^{+12}_{-4}$ keV, and $kT_C = 5.2^{+1.4}_{-1.0}$ keV, respectively.
Since systematic uncertainties in the BGD level and in the response
have little influence on the estimated temperatures, it is certain
that A115 has a significant temperature structure along the main
to subcluster path. This cluster is, therefore, probably undergoing an
early phase of a subcluster merger.

When the merging scenario is assumed, a relative colliding velocity
$v_{col}$ of the subcluster necessary to heat up the ICM temperature
$kT$ from $\sim 5$ keV to $\sim 11$ keV\@.  We further assume that the
two subclusters are to cause a head-on collision and that their
kinetic energies are completely converted to thermal energy.  Thus,
necessary value of $v_{col}$ to result in a temperature rise of $k
\Delta T = 6$ keV is given by,
\begin{eqnarray*}
v_{col} & = & \left ( \frac{3 k \Delta T}{\mu m_p} \right )^{1/2} \\
& \simeq & 1700 \quad [{\rm km\ s}^{-1}]
\end{eqnarray*}
where $\mu$ is the mean molecular weight (0.6) in amu, and $m_p$ is
the proton mass, respectively.  Beers {\it et al.}\ (1983) calculated
a velocity dispersion for 6 galaxies which were unambiguously
associated with A115 based on their mapping observation in the R
band, and obtained $\sigma_r \simeq 1200$ km s$^{-1}$. The resulting
$\beta (= \mu m_p \sigma_r^2/(kT))$ for $kT_B \sim 11$ keV turns out
to be $\sim 0.8$, which is reasonable.  In the cooling flow picture,
the temperature of the hot-phase gas is obtained to be $\sim 6$ keV,
which results in a somewhat low $\beta$ value.

Taking, $L_X \sim 5.1 \times 10^{44}$ erg s$^{-1}$ and $V \sim
9.8$ Mpc$^3$, in the spherical region B with $2'$ radius, we obtain
the average gas density in this region to be $n_{B} \simeq 3.8
\times 10^{-4}$ cm$^{-3}$.  The cooling time of the ICM in this region
is estimated as
\[
t_{\rm cool}  \simeq  2.2 \times 10^{11} \ {\rm [yr]}
\left( \frac{n_{B}}{3.8 \times 10^{-4}\ [{\rm cm}^{-3}] } 
\right)^{-1}
\left( \frac{kT_B}{11\ {\rm [keV]}} \right) ^{1/2}.
\]
Therefore, cooling is not important in the region B\@. As a result,
heating due to the collision of subclusters would effectively raise
the ICM temperature.

Above considerations suggest that the shock heating caused by a
subcluster merger is more important than the cooling flows in creating
the observed spatial variation of temperature.  {\it ASCA} and {\it
ROSAT} observations detect spatial temperature structures from several
clusters which are likely to be caused by subcluster mergers.
For example, it became clear with {\it ASCA} observations of the Coma
cluster that there are a low temperature region ($\sim 5$ keV) in the
northeast and a high temperature one ($\sim 12$ keV) in the southwest
(Honda {\it et al.}\ 1996). This provides evidence of a merger in this
cluster.
Many other clusters show significant temperature structures, such as
A3558 (Markevitch \& Vikhlinin 1997), Triangulum Australis cluster
(Markevitch {\it et al.}\ 1996a), A2319, A2163, A665 (Markevitch
1996b), A2256 (Briel \& Henry 1994, Roettiger {\it et al.}\ 1995,
Markevitch 1996b), A2163 (Markevitch {\it et al.}\ 1994, 1996c), and
A754 (Henry \& Briel 1995, Henriksen \& Markevitch 1996).

We cannot sufficiently constrain metal abundance and its variation in
this cluster because of poor statistics.  The best-fit values of
abundance in the spectral analysis are 0.2, 0.1 and 0.2 solar for the
regions A, B and C, respectively.  The region B shows a somewhat lower
value.  Since the abundance distribution probably reflects the local
density of galaxies (Ezawa {\it et al.}\ 1997), it may be reasonable
that the linking region with low galaxy density shows a low
metallicity.
%We can also infer that the ICM in this
%region is probably not yet mixed due to the merger.

\section{Conclusion}

{\it ASCA} observations of the cluster of galaxies A115 have revealed
significant temperature variation in the ICM for the first time.  The
hardness ratio (H.R.) profile shows that the linking region between
the two X-ray peaks has the highest temperature, which is confirmed
with the spectral analysis. To investigate the origin of this
temperature variation, we have looked into two possible scenarios:
cooling flows in the main cluster and a shock heating due to the
collision of the subcluster.

The {\it ASCA} spectrum of the main cluster is consistent with a
cooling flow model with the mass deposition rate lower than the {\it
Einstein} result. Our upper limit on the mass-flow rate is $419
M_{\odot}$ yr$^{-1}$.  The simulated H.R. profile including the
cooling flow ($313 M_{\odot}$ yr$^{-1}$) can reproduce the
H.R. profile around the main cluster.  However, the cooling time in
the center of the main cluster is estimated to be $t_{\rm cool} \sim
1.7 \times 10^{10}$ yr, which is comparable to the Hubble time. The
cooling flow may not be the dominant process causing the temperature
variation in A115.

On the other hand, merger scenario can explain the high temperature
region in terms of a collision between the two subclusters, as
suggested from its morphology.  Temperatures are estimated by
correcting for the contaminating flux due to the PSF of the XRT as
$\sim 5$ keV and $\sim 11$ keV, for subcluster centers and the linking
region, respectively.  Observed velocity dispersion of galaxies in
A115 and the ICM temperature of 5 keV indicates $\beta_{\rm spec}$ to
be $\sim 0.8$, which is within a reasonable range for clusters.
Cooling time of the gas in the linking region is much longer than the
Hubble time because of the low density. Based on these features, we
conclude that heating due to the collision of the subcluster is
probably the major cause of the temperature variation in A115. It is
also likely that the presence of cooing flows in the main cluster
makes the temperature distribution more complicated.

\bigskip
We thank anonymous referee for pointing out the possibility of cooling
flows. We would like to express our thanks to the {\it ASCA} team and
the ASCA\_ANL, SimASCA, and SimARF software development team for
constructing excellent frameworks for the analysis.  We also thank
F. Makino and S. Uno for valuable advice on the analysis method.
R. S. acknowledges support from the Japan Society for the Promotion of
Science for Young Scientists.

\clearpage

%%% Table 1 %%%%%%%%%%%%%%%%%%%%%%%%%%%%%%%%%%%%%%%%%%%%%%%%%%%%%%%%%%%%
\begin{table}[htb]
%\begin{center}
\caption{{\it ASCA } observation of A115.}
\vspace{3mm}
\begin{tabular}{ccccc} \hline \hline  
Detector & DP Mode & Exposure Time$^a$ & Counting Rate$^b$ 
& Flux(0.5-10keV)$^c$\\ 
& &      [s]& [counts s$^{-1}$] & [erg cm$^{-2}$ 
s$^{-1}$] \\
\hline 
SIS-0 &     & & 0.291  &\\[-3mm] 
      & Faint & 32,000 &&  (9.6 $\pm 0.4) \times 10^{-12}$ \\[-3mm]
SIS-1 &     & &  0.223 &\\
GIS-2 &  && 0.183  &\\[-3mm] 
& PH nominal & 33,100 && (1.1 $\pm 0.1) \times 10^{-11}$ \\[-3mm] 
GIS-3 &  && 0.220 &\\ 
\hline \hline
\multicolumn{5}{p{135mm}}{{\small $a$ : 
Average for SIS0 and SIS1, and for GIS2 and GIS3.}} \\
\multicolumn{5}{p{135mm}}{{\small $b$ : 
Integration regions are $11' \times 11'$ for SIS and 10$'$ radius for GIS.
}} \\
\multicolumn{5}{p{135mm}}{{\small $c$ :
Assuming a single temperature Raymond-Smith model with a best fit
temperature $kT$ = 5 keV ($z$ = 0.1971). The errors indicate 90\%
statistical ones.
}} \\
\end{tabular}
%\end{center}
\label{table_1}
\end{table}

%%% Table 2 %%%%%%%%%%%%%%%%%%%%%%%%%%%%%%%%%%%%%%%%%%%%%%%%%%%%%%%%%%%%

\begin{table}[htb]
\caption{The number of photons in each integration region}
\label{table_2}
\vspace{3mm}
\begin{tabular}{cp{2.4cm}p{2.4cm}p{2.4cm}} \hline \hline  
Integration Radius~ & \multicolumn{3}{c}{Counting Rate$^a$ [$\times$ 10$^{-2}$ 
counts s$^{-1}$]} \\
$[$arcmin.](sensor) & Region A & Region B & Region C \\ \hline
&&& \\[-3mm]
2  (GIS) & 4.98 (0.23) & 3.73 (0.22) & 2.83 (0.21) \\
2  (SIS) & 8.89 (0.54) & 6.61 (0.54) & 4.50 (0.44) \\
&&& \\[-3mm]
3  (GIS) & 8.31 (0.51) & 8.16 (0.51) & 5.29 (0.47) \\
3  (SIS) & 14.6 (1.17) & 14.6 (1.13) & 9.54 (1.05) \\
&&& \\[-3mm]
5  (GIS) & 14.9 (1.38) & & \\
10 (GIS) & 25.5 (5.43) & & \\
&&& \\[-3mm]
\hline \hline 
\multicolumn{4}{p{9.5cm}}{{\small a: Counting rates in round brackets
indicate those of background data. }}
\end{tabular}
\end{table}

%%% Table 3 %%%%%%%%%%%%%%%%%%%%%%%%%%%%%%%%%%%%%%%%%%%%%%%%%%%%%%%%%%%%
\begin{table}[htb]
\caption{The fitting parameters for each
integration region. The GIS and SIS data are jointly fitted.}
\label{table_3}
\vspace{3mm}
\begin{tabular}{cccccc} \hline \hline 
Integration &  Integration & \multicolumn{4}{c}{Spectral Model $^{a}$} \\
Region  & Radius & \multicolumn{4}{c}{ Raymond \& Smith } \\
     &\footnotesize{[arcmin.]} & ~~~$kT$[keV]~~~ & ~~~Abundance~~~ &
~~~$N_H$[$\times$ 10$^{21}$cm$^{-2}$]~~~ & ~~~$\chi^2/\nu$~~~ \\ \hline
&&&&& \\[-4mm]
A & 2 & 4.85$^{+0.45}_{-0.40}$ & 0.21$^{+0.08}_{-0.08}$ 
& 0.91$^{+0.28}_{-0.25}$ & 163.8/159\\
A & 3 & 5.37$^{+0.42}_{-0.39}$ & 0.15$^{+0.07}_{-0.07}$ 
& 0.76$^{+0.22}_{-0.21}$ & 194.0/159\\
A & 5~$^*$  & 5.64$^{+0.68}_{-0.53}$ & 0.14$^{+0.09}_{-0.08}$ 
& 0.66$^{+0.46}_{-0.46}$ & 96.29/79\\
A & 10~$^*$ & 5.05$^{+0.47}_{-0.43}$ & 0.28$^{+0.09}_{-0.09}$ 
& 0.96$^{+0.41}_{-0.41}$ & 88.86/79\\
B & 2 & 6.25$^{+0.93}_{-0.77}$ & 0.10$^{+0.10}_{-0.10}$
& 0.84$^{+0.37}_{-0.33}$ & 215.7/159 \\
B & 3 & 5.74$^{+0.54}_{-0.42}$ & 0.13$^{+0.07}_{-0.06}$ 
& 0.93$^{+0.23}_{-0.23}$ & 192.1/159 \\
C & 2 & 5.08$^{+0.71}_{-0.61}$ & 0.21$^{+0.12}_{-0.12}$ 
& 1.32$^{+0.44}_{-0.41}$ & 144.2/159\\
C & 3 & 5.52$^{+0.62}_{-0.52}$ & 0.21$^{+0.09}_{-0.09}$ 
& 1.06$^{+0.31}_{-0.31}$ & 159.6/159\\
&&&&& \\[-3mm] \hline \hline
\multicolumn{6}{l}{{\small $a$ : $N_H$ = free, redshift = 0.1971 (fixed).
The errors indicate 90\% statistical ones.}} \\
\multicolumn{6}{l}{{\small $*$ : 
Only GIS data are used because the regions are not covered 
with the SIS FOV.}} \\
\end{tabular}
\end{table}

%%% Table 1.5 %%%%%%%%%%%%%%%%%%%%%%%%%%%%%%%%%%%%%%%%%%%%%%%%%%%%%%%%%%%%
\begin{table}[htb]
\caption{The results of spectral fit to inspecting the 
cooling flow in the region A within 2$'$ radius .  The GIS and SIS
data are jointly used.}
\label{table_cf}
\vspace{3mm}
\begin{tabular}{ccccccc} \hline \hline 
\multicolumn{2}{c}{ICM}          & & \multicolumn{2}{c}{cool component} &\\
\multicolumn{2}{c}{Raymond-Smith model} &
& \multicolumn{2}{c}{cooling flow model $^a$} &\\
kT[keV] & Abundance &~~&  slope & 
mass-flow rate [M$_{\odot}$] &~~& $\chi^2/\nu$ \\ \hline
4.85 $^{+0.45}_{-0.40}$ & 0.21 $^{+0.08}_{-0.08}$ & & 
--- & --- && 163.8/159 \\ 
&&&&&& \\[-3mm]
5.45 $^{+0.39}_{-0.37}$ & 0.17 $^{+0.09}_{-0.09}$ & & 
$-$0.5 (fixed)& 313 (fixed) && 170.5/160 \\ 
5.63 $^{+0.43}_{-0.42}$ & 0.18 $^{+0.09}_{-0.09}$ & & 
0.0 (fixed)& 313 (fixed) && 168.5/160 \\ 
6.53 $^{+0.82}_{-0.70}$ & 0.19 $^{+0.09}_{-0.09}$ & & 
0.7 (fixed)& 313 (fixed) && 168.2/160 \\ 
&&&&&& \\[-3mm]
4.85 $^{+0.45}_{-0.29}$ & 0.21 $^{+0.08}_{-0.08}$ & & 
$-$0.5 (fixed)& $<$ 282 && 163.8/159 \\ 
4.85 $^{+0.64}_{-0.29}$ & 0.21 $^{+0.09}_{-0.08}$ & & 
0.0 (fixed)& $<$ 362 && 163.8/159 \\ 
5.00 $^{+0.95}_{-0.44}$ & 0.21 $^{+0.09}_{-0.08}$ & & 
0.58 (fixed)& 89 $^{+330}_{-89}$ && 163.6/159 \\ 
5.01 $^{+0.94}_{-0.45}$ & 0.21 $^{+0.07}_{-0.08}$ & & 
0.7 (fixed)& 88 $^{+319}_{-88}$ && 163.6/159 \\ \hline \hline
\multicolumn{7}{p{13cm}}{{\small $a$ : We adopt the model made by 
Mushotzky \& Szymkowiak (1988).  Low temperature is fixed with 0.1
keV, and high temperature is linked up with the temperature of ICM.
Redshift and column density are also fixed with 0.1971 and galactic
value, respectively.  The errors indicate 90\% statistical ones.}} \\
\end{tabular}
\end{table}

%%% Table 4 %%%%%%%%%%%%%%%%%%%%%%%%%%%%%%%%%%%%%%%%%%%%%%%%%%%%%%%%%%%%
\begin{table}[htb]
\caption{The ICM temperature in each integration region using only SIS
data, which have less contamination than the GIS data.  The used
responses are calculated assuming an isothermal ICM throughout the
whole cluster (``Raw Values'' in the table), and a non isothermal
ICM which is corrected for the flux contamination due to the XRT
(``Real Values'' in the table), respectively.  }
\label{table_4}
\vspace{3mm}
\begin{tabular}{cccccc} \hline \hline 
&&  \multicolumn{3}{c}{Temperature [keV]} & Reference \\ 
&& Region A & Region B & Region C  & Section \\ \hline
Raw values &&
5.0$^{+0.6}_{-0.6}$ & 6.8$^{+1.6}_{-1.1}$ & 5.3$^{+1.2}_{-0.9}$ & 
\ref{sec-spec-all} \\
Real Values &&
4.9$^{+0.7}_{-0.6}$ & 11$^{+12}_{-4}$ & 5.2$^{+1.4}_{-1.0}$ & 
\ref{sec-real-temp} \\
\hline \hline
\multicolumn{6}{l}{{\small NOTE : 
The errors indicate 90\% statistical ones.}} \\
\end{tabular}
\end{table}

%%% Figure 1 %%%%%%%%%%%%%%%%%%%%%%%%%%%%%%%%%%%%%%%%%%%%%%%%%%%%%%%%%%%

%%% Figure 2 %%%%%%%%%%%%%%%%%%%%%%%%%%%%%%%%%%%%%%%%%%%%%%%%%%%%%%%%%%%

%%% Figure 3 %%%%%%%%%%%%%%%%%%%%%%%%%%%%%%%%%%%%%%%%%%%%%%%%%%%%%%%%%%%

%%% Figure 4 %%%%%%%%%%%%%%%%%%%%%%%%%%%%%%%%%%%%%%%%%%%%%%%%%%%%%%%%%%%

%%% Figure 5 %%%%%%%%%%%%%%%%%%%%%%%%%%%%%%%%%%%%%%%%%%%%%%%%%%%%%%%%%%%

%%% Figure cf_hr %%%%%%%%%%%%%%%%%%%%%%%%%%%%%%%%%%%%%%%%%%%%%%%%%%%%%%%%%%%

%%% Figure 6 %%%%%%%%%%%%%%%%%%%%%%%%%%%%%%%%%%%%%%%%%%%%%%%%%%%%%%%%%%%

\end{document}